\documentclass{aa}  

\usepackage{graphicx}
\usepackage[normalem]{ulem}
\usepackage{orcidlink} 
\usepackage{hyperref}
\hypersetup{
    colorlinks=true,
    allcolors=blue}
\usepackage{booktabs}
\usepackage{comment}
\usepackage{txfonts}
\usepackage{caption}
\usepackage{subcaption}
\usepackage[flushleft]{threeparttable}
\usepackage{hyperref}
\usepackage{dirtytalk}
\usepackage{titlesec}
\begin{document}

   \title{Discovery of a New Transient X-ray Source in M51 Detected in the Ultraluminous Regime}

    \author{A. Yilmaz
          \inst{1}\thanks{\email{anastasiya.yilmaz@inaf.it}}\orcidlink{0000-0003-1133-1684}
          \and
          D. Lan\v{c}ov\'{a}\inst{2,3}\thanks{\email{debora.lancova@physics.cz}}\orcidlink{0000-0003-0826-9787}
          \and 
          F. F\"urst\inst{4}\orcidlink{0000-0003-0388-0560}
          \and
          D. Calder\'on\inst{5}\orcidlink{0000-0002-9019-9951}
          \and
          S. Reyero Serantes\inst{6,7}\orcidlink{0000-0002-3155-7237}
          \and
          E. Bozzo\inst{8}\orcidlink{0000-0002-8201-1525}
          \and
          G. Piano\inst{1}\orcidlink{0000-0002-9332-5319}
          \and
          \\J. Svoboda\inst{9}\orcidlink{0000-0003-2931-0742}
          \and
          C. Pittori\inst{10}\orcidlink{0000-0001-6661-9779}
          \and
          M. Tavani\inst{1}\orcidlink{0000-0003-2893-1459}}

            \institute{INAF – Istituto di Astrofisica e Planetologia Spaziali, Via del Fosso del Cavaliere 100, I-00133 Roma, Italy
             \and
             Nicolaus Copernicus Astronomical Centre, Polish Academy of Sciences, Bartycka 18, 00-716 Warsaw, Poland
             \and
             Research Centre for Computational Physics and Data Processing, Institute of Physics, Silesian University in Opava, Bezru\v{c}ovo n\'am.~13, CZ-746\,01 Opava, Czech Republic
             \and
             European Space Agency (ESA), European Space Astronomy Centre (ESAC), Camino Bajo del Castillo s/n, 28692 Villanueva de la Cañada, Madrid, Spain
             \and
             Max-Planck-Institut für Astrophysik, Karl-Schwarzschild-Straße 1, 85748 Garching, Germany
             \and
             Institut für Physik und Astronomie, Universität Potsdam, Karl-Liebknecht-Str. 24/25, 14476 Potsdam, Germany
             \and 
             Centro de Astrobiolog\'ia (CAB, CSIC-INTA), 28692 ESAC Campus, Villanueva de la Cañada, Madrid, Spain
             \and
             Department of Astronomy, University of Geneva, Chemin d’Ecogia 16, CH-1290 Versoix, Switzerland
             \and
             Astronomical Institute of the Czech Academy of Sciences, Bo\v{c}n\'\i -II 1401, Praha 4 Prague, 141~00, Czech Republic
             \and
             INAF - Osservatorio Astronomico di Roma, I-00078 Monte Porzio Catone, Italy
             }
   \date{Received ; accepted }

 \abstract{
We report the discovery of a~transient X-ray source in the nearby galaxy M51, detected with Swift, XMM-Newton, and Chandra between June and August 2018. 
The source, M51 X* (RA = 13:29:45.46, Dec = +47:09:32.48; J2000), exhibits strong variability over the available archival coverage spanning more than two decades. During its 2018 period of activity, the source reached a luminosity of $\sim 7.8 \times 10^{38}\,\mathrm{erg\,s^{-1}}$ in the XMM-Newton observations. Subsequently, Swift-XRT detected a considerably brighter phase, placing the source firmly in the ultraluminous regime. The luminosity then declined rapidly, with the final Chandra detection at the end of August showing the source at $\sim 8 \times 10^{37}\,\mathrm{erg\,s^{-1}}$. The XMM-Newton spectra are consistently soft, while the limited Swift/XRT photon statistics do not allow us to constrain possible spectral changes.No significant pulsations were detected in the timing analysis. Together with the large long-term variability and the recurrence of the X-ray activity, these properties suggest that M51 X* hosts a stellar-mass compact object undergoing episodes of super-Eddington accretion. The nature of the accretor remains uncertain, but the observed spectral and temporal properties suggest that M51 X* harbors a~stellar-mass compact object reaching accretion rates exceeding the Eddington limit for a stellar-mass accretor.

}
\maketitle

\section{Introduction}

\renewcommand{\sectionautorefname}{Section}

Ultraluminous X-ray sources (ULXs) are compact, point-like sources whose characteristic luminosities exceed the Eddington limit, usually observed in nearby galaxies \citep[see][for a~review]{Kaaret17, Fabrika21,King23}. Since their discovery, they have been proposed as hosts of intermediate-mass black holes (10$^2$-10$^5$~\text{M}$_\odot$) as such masses could explain the extreme luminosities \citep{col99,kon+04,mil+04,liu08}. However, this view was challenged by the discovery of coherent pulsations in some of those systems, such as in the M82 X-2 \citep{bac+14}. This fact confirmed the presence of accreting neutron stars in some ULXs. These are known as pulsating ultra-luminous X-ray sources (PULXs), which are thought to host magnetized neutron stars accreting at super-Eddington rates. Despite the detections of more PULXs over the last decade \citep{fur+16, isr+17a, isr+17b, carpano18, sath19, rod20}, the physical mechanism responsible for their luminosities remains unknown.
  
   \begin{figure*}
        \includegraphics[width=0.98\textwidth]{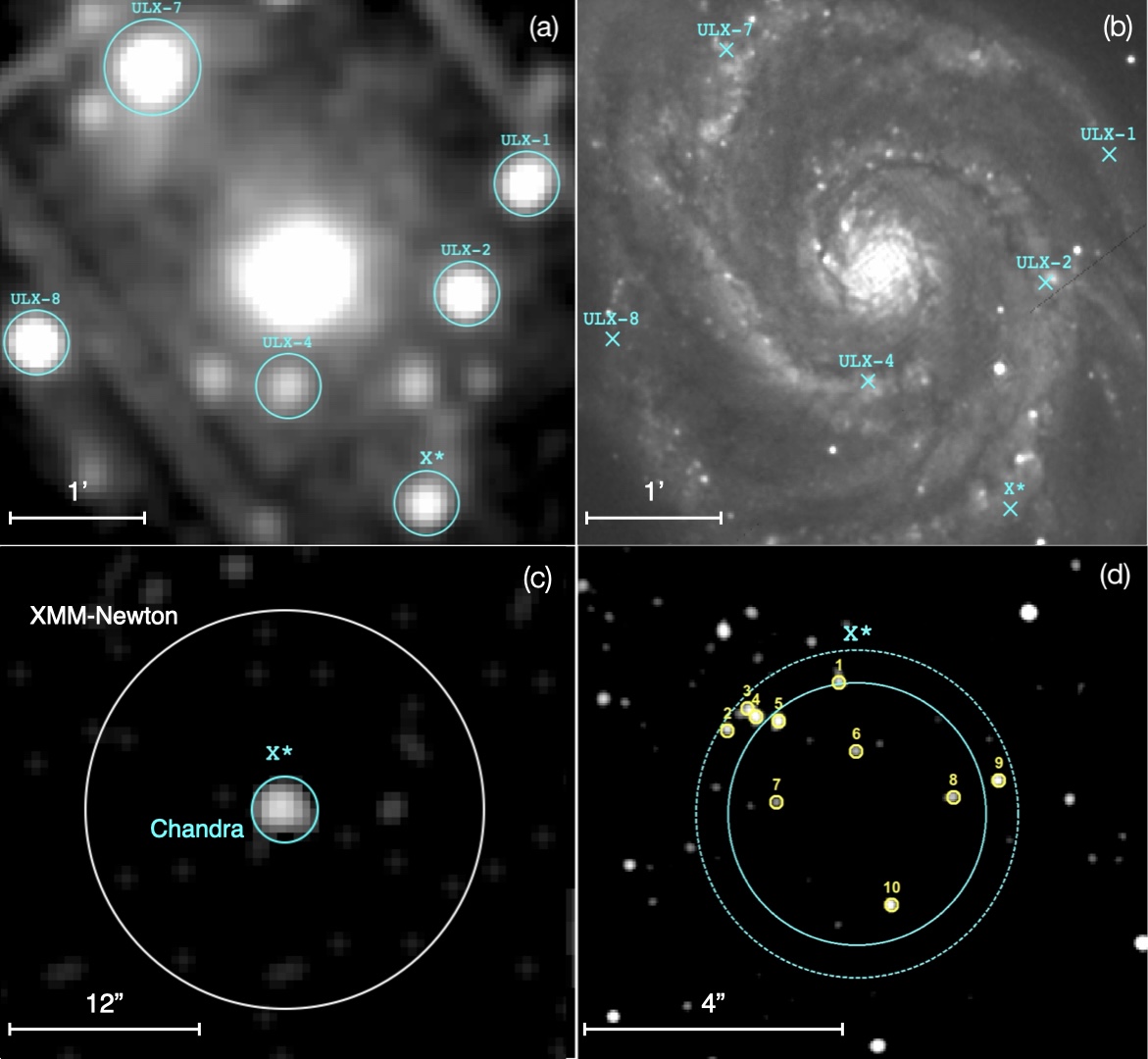}
        \includegraphics[width=0.98\textwidth]{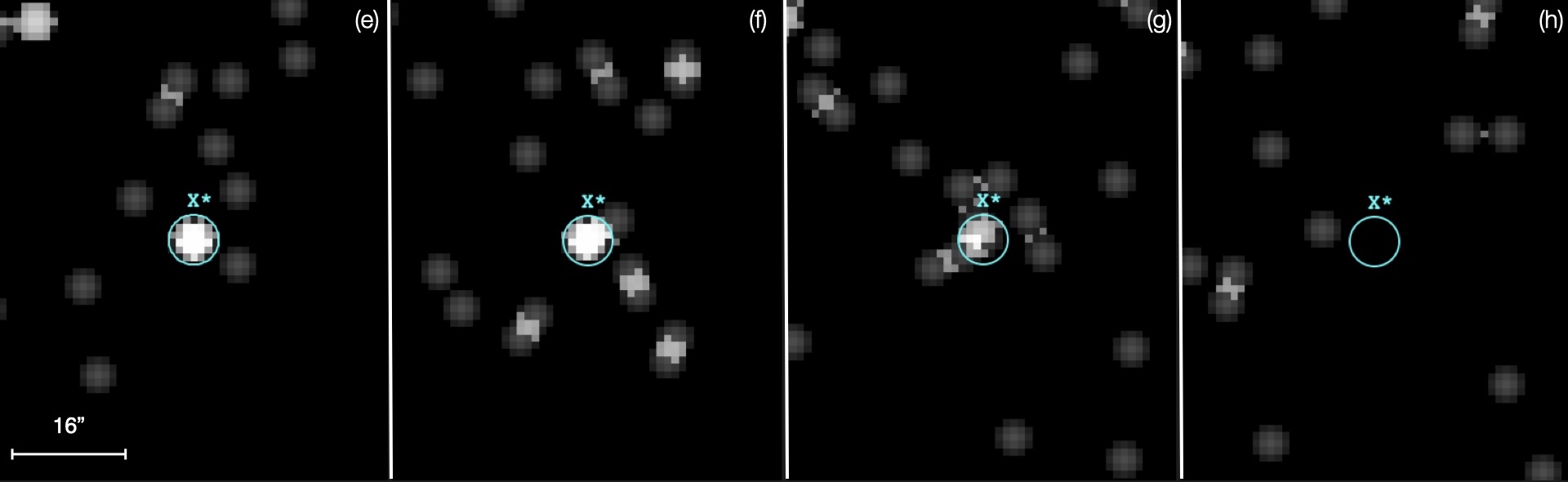}
        \caption{(a) XMM-Newton EPIC-PN image of M51 with highlighted known ULXs, (b) XMM-Newton OM image of M51 in the B-band obtained from X2. (c) Chandra image of M51 X*. The large white circle marks the XMM-Newton extraction region (d) Hubble Space Telescope (HST) WFC3 image in the F275W filter of the same $2\arcsec$ region. (e)-(f) Swift-XRT images of M51 X* obtained from S2 (divided into separate observations), (g) S3 and (h) S4.} The blue dashed circle corresponds to the same aperture when considering an error margin of ($0.5\arcsec$ radius), while the blue circle indicates the Chandra aperture with a~radius of $2\arcsec$. The yellow circles indicate all possible counterparts of the X-ray source in the ultraviolet (UV) branch; see text for details.  The new X-ray transient is marked as X*, located at 13:29:45.46 +47:09:32.48 (J2000).\label{fig:img}
    \end{figure*}

\begin{table*}
\centering
\caption{\label{table:obs} XMM-Newton, Chandra, Swift, and AGILE observations of M51 used in the discussion of this paper (in chronological order).}
\renewcommand{\arraystretch}{1.2}
\begin{threeparttable}
\resizebox{\textwidth}{!}{%
\begin{tabular}{@{}cccccccc@{}}
\hline
\hline
DataID \tnote{a} & ObsID & Exposure & Total Counts & Date & ${\rm F_{\rm x}}$ & ${\rm L_x}$ & Energy Range \\
& & (ks) & & & (${\rm erg\ s^{-1}\ cm^{-2}}$) & (${\rm erg\ s^{-1}}$) & \\
\hline
X1\tnote{*} & 0303420101 & 54.11 & --& 2006-05-20 & $17.9 \times 10^{-15}$ & $1.6 \times 10^{38}$ & 0.5--7.0 keV \\
C1\tnote{*} & 13815 & 67.18 & -- & 2012-09-23 & $4.7 \times 10^{-17}$ & $4.2 \times 10^{35}$ & 0.5--7.0 keV \\
S1 & 00032017092 & 1.63 & 3 & 2018-06-12 & $3.4 \times 10^{-14}$ & $3.0 \times 10^{38}$ & 0.5--7.0 keV \\
X2 & 0830191501 & 63.00 & 1082 & 2018-06-13 & $79.9^{+3.5}_{-3.3} \times 10^{-15}$ & $7.0^{+0.3}_{-0.3} \times 10^{38}$ & 0.5--7.0 keV \\
X3 & 0830191601 & 63.00 & 1320 & 2018-06-15 & $88.1^{+5.0}_{-5.2} \times 10^{-15}$ & $7.8^{+0.4}_{-0.4} \times 10^{38}$ & 0.5--7.0 keV \\
S2 & 00032017108/109 & 3.8 & 11 & 2018-08-14/18 & $5.9^{+4.0}_{-4.2} \times 10^{-12}$ & $5.2^{+3.5}_{-3.7} \times 10^{40}$ & 0.5--7.0 keV\\
S3\tnote{*} & 00032017110 & 2.0 & -- & 2018-08-20 & $4.7 \times 10^{-12}$ & $4.1 \times 10^{40}$ & 0.5--7.0 keV \\
S4\tnote{*} & 00032017111 & 1.8 & -- & 2018-08-23 & -- & --&0.5--7.0 keV \\

C2 & 20998 & 19.82 & 18 & 2018-08-31 & $9.5^{+3.8}_{-5.1} \times 10^{-15}$ & $0.8^{+0.3}_{-0.4} \times 10^{38}$ & 0.5--7.0 keV \\
X4\tnote{*} & 0852030101 & 77.00 & -- & 2019-07-11 & $15.3 \times 10^{-15}$ & $1.3 \times 10^{38}$ & 0.5--7.0 keV \\
\hline
& & & AGILE & & & & \\
\hline
A\tnote{*} & -- & -- & -- & 2018-06-12/08-31 & $8.48 \times 10^{-11}$ & -- & 100 MeV--50 GeV \\
\hline
\hline
\end{tabular}%
}
\begin{tablenotes}
\item[*] Upper limit
\item[a] A, C, S and X denote AGILE, Chandra, Swift and XMM-Newton observations, respectively, and the numbering follows their chronological order.
\end{tablenotes}
\end{threeparttable}
\end{table*}

The persistent nature of a~significant portion of the known ULX population is often attributed to the presence of a~high-mass companion star, with the accretion disk being fed by stellar winds \citep{Martinez17, elmel+19}. However, this alone cannot fully account for the observed super-Eddington accretion rates or the long-term variability. Transient ULXs, which display large luminosity variations over short timescales, present an even greater challenge to accretion models \citep{Dage21}. The transient behavior observed in many ULXs is hard to characterize into definitive groups, as the observed luminosities and short/long-term variability show no apparent pattern.

The observed super-Eddington luminosities in ULXs can be interpreted through various theoretical scenarios, for example:
(a) accreting stellar-mass black holes, where density perturbations within the disk might overcome the photon trapping. As a~result, the diffusion timescale increases relative to the accretion timescale, which allows the escape of more photons from the inner regions of the disk, leading to super-Eddington luminosities \citep{Arons92,Gammie98,Begelman02}; 
(b) relativistic beaming in jets along the line of sight \citep{Falcke99, Markoff}. This jet-disk model was found to be consistent with the luminosity function of X-ray binaries (XRBs) in nearby galaxies. However, this still requires a~black hole mass of $\sim 1000~\text{M}_\odot$ in the case of ULXs, assuming an isotropic disk model \citep{Kording02}; 
(c) a~decrease in the electron opacity for X-rays in accreting neutron stars with magnetic fields of $\sim \rm 10^{12}-10^{13}~G$ \citep{Canuto71,Mushtukov15};
(d) geometrical beaming due to the funnel-shaped structure of thick, radiation-pressure-dominated accretion disks. In principle, these can lead to an anisotropic emission pattern that boosts the apparent luminosity along the line of sight \citep{kin+01, kin09}. In this scenario, moderate beaming factors could account for ULX luminosities for stellar-mass accretors, including neutron stars.

The interacting galaxy system M51, located at a~distance of 8.58$ \pm{0.10}$~Mpc \citep{mcq16} hosts a~large population of XRBs  \citep[334 candidates, ][]{Rice21,Sanatombi23}. At least 9 ULXs with luminosities exceeding $ \rm 10^{39}~erg\ s^{-1}$ were identified to be persistent \citep{ter04}. M51 consists of the spiral galaxy M51a (also known as the Whirlpool Galaxy) and the dwarf galaxy M51b, also referred to as NGC 5194 and NGC 5195, respectively. Due to its favorable face-on orientation in the sky and the expansive number of observations, M51 hosts one of the most studied ULX populations in a~single galaxy.  

In this paper, we report the discovery of a new transient ultraluminous X-ray source in M51 (hereafter M51 X*), using  Neil Gehrels Swift Observatory  \citep[Swift, ][]{Gehrels2004}, X-ray Multi-Mirror Mission \citep[XMM-Newton, ][]{Jansen01}, and Chandra X-ray Observatory \citep[Chandra, ][]{Weisskopf02} observations. During its 2018 period of activity, the source was first detected in June, when it reached an X-ray luminosity just below the canonical ULX threshold. Approximately two months later, Swift-XRT observations revealed a significantly brighter phase, during which the source reached the ultraluminous regime. The available observations do not allow us to determine whether these detections correspond to a single outburst or to separate episodes.
The spectral and long-term evolution of the source does not follow the canonical behavior observed in typical black hole or neutron star X-ray binaries, indicating that M51 X* represents a~rare and extreme accretion regime, consistent with its classification as a~candidate ULX. 

This work is organized as follows: in \autoref{observations}, we describe the observations, data reduction procedures, and analysis methods. The main observational results are presented in \autoref{results}. In \autoref{discussion}, we discuss the nature of the source and explore possible physical scenarios that could explain its transient behavior and high luminosity. Our conclusions are summarized in \autoref{conclusions}.

\begin{figure*}[h!]
\centering
\hspace{-0.5cm}
        \includegraphics[width=\textwidth]{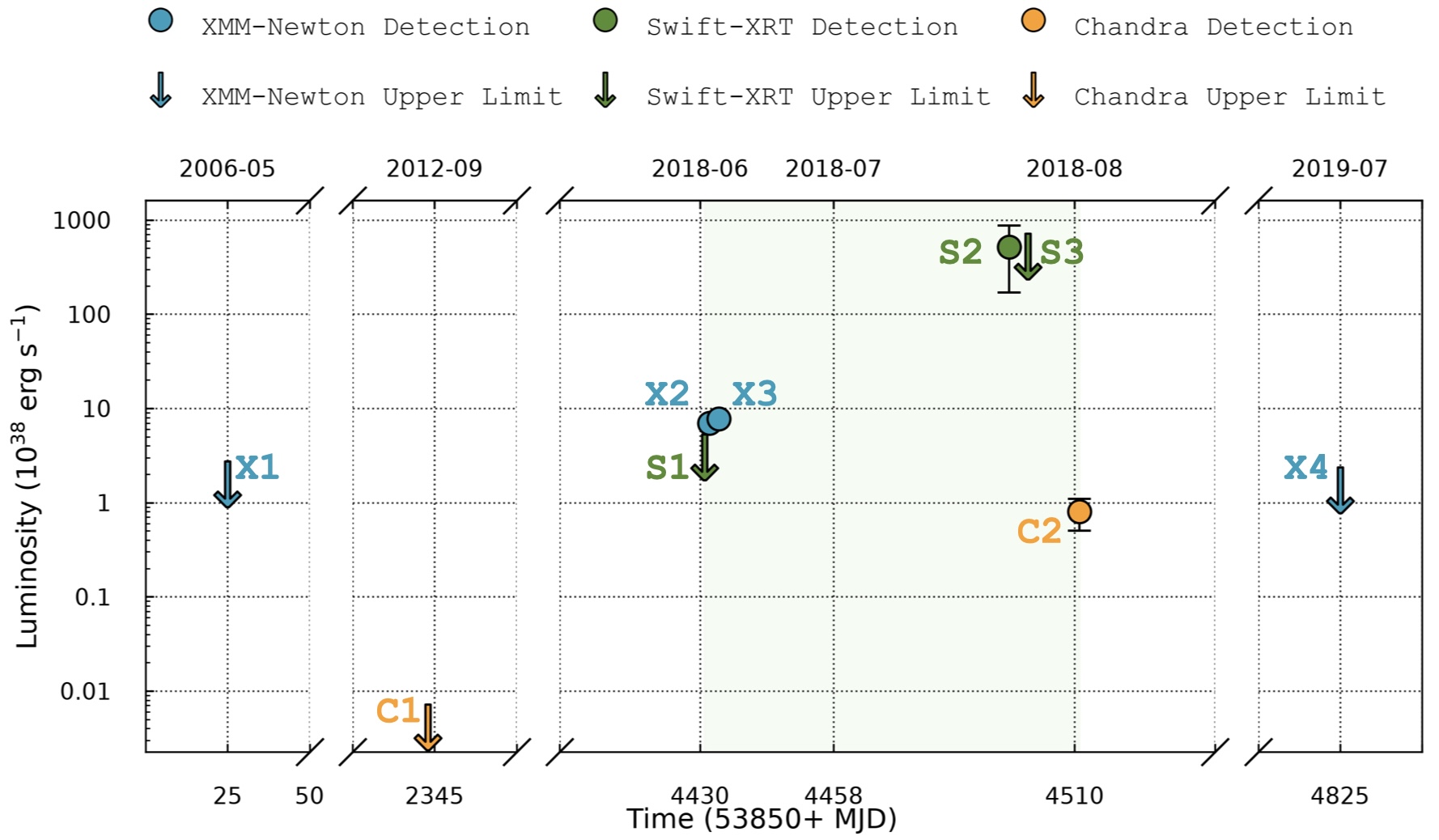}
        \caption{Long-term lightcurve of M51 X* based on archival XMM-Newton, Swift-XRT, and Chandra observations listed in \autoref{table:obs}. The shaded region marks the period during which the source was in outburst, downward arrows denote upper limits and full circles are used to highlight the detections. Blue, orange and green represent XMM-Newton, Chandra and Swift-XRT observations, respectively. 
        }\label{fig:lcurve}
    \end{figure*}

\section{Observations and Data Reduction}\label{observations}

M51 was extensively observed in soft X-rays by Swift, Chandra, and XMM-Newton between 2000 and 2024, enabling a~long-term study of the ULX population as more sources had been identified in the galaxy \citep{Sanatombi23,Imbrogno24}. In this section, we describe only the observations used to study the outburst and presented in the results. In \autoref{table:obs}, we list only the 
archival AGILE, Chandra, Swift, and XMM-Newton observations used for detection and further analysis, separating the observations where M51 X* was detected and observations used to obtain upper limits before and after the detection.

The source was observed dormant or below the detection threshold in all XMM-Newton, Chandra, and Swift observations from 2000 until the first detection in the 1.63-ks Swift X-Ray Telescope (Swift-XRT) observation (ObsID:00032017092) on June 12, 2018. Shortly after the first detection with Swift, the flux increased significantly during two XMM-Newton observations (ObsIDs: 0830191501, X2, and 0830191601, X3). Unfortunately, these coincided with the end of the month-long observing campaign, with five evenly distributed observations. The Chandra observation (ObsID: 20998, C2), subsequent to the XMM-Newton detections, revealed a~significant drop in the flux by almost an order of magnitude. The remainder of the Chandra observations either were not long enough for a~proper detection or the source was outside the field of view.

\subsection{Chandra X-ray Observatory}

We reduced the raw Chandra data to obtain cleaned level 2 event files using the \texttt{chandra\_repro} command within the Chandra Interactive Analysis of Observations software \citep[\texttt{CIAO} v.4.15,][]{Ciao}, with calibration files v.4.10.4. The source was detected only in C2, with a total of 18 source counts. Given the low count statistics, we grouped the C2 spectrum using \texttt{ftgrouppha} with optimal binning and analyzed it using BXA and C-statistics.

We also used C2 to obtain a more precise source position. We extracted a full-detector image together with the corresponding exposure and point-spread-function (PSF) maps in the $0.5$--$7.0$~keV energy band and performed source detection with \texttt{wavdetect}. The same procedure was applied to the remaining Chandra observations in which the source was within the field of view. For observations without a significant detection, we used \texttt{srcflux} to derive flux upper limits. The deepest constraint was obtained from the 67.18~ks C1 observation, with $F_{\rm x}<4.69\times10^{-17}~\mathrm{erg\,s^{-1}\,cm^{-2}}$ in the $0.5$--$7.0$~keV band, corresponding to $L_{\rm x}<4.13\times10^{35}~\mathrm{erg\,s^{-1}}$. The C1 upper limit and C2 detection are listed in \autoref{table:obs}.

\begin{figure*}[h!]
\centering
\hspace{-0.5cm}
        \includegraphics[width=\textwidth]{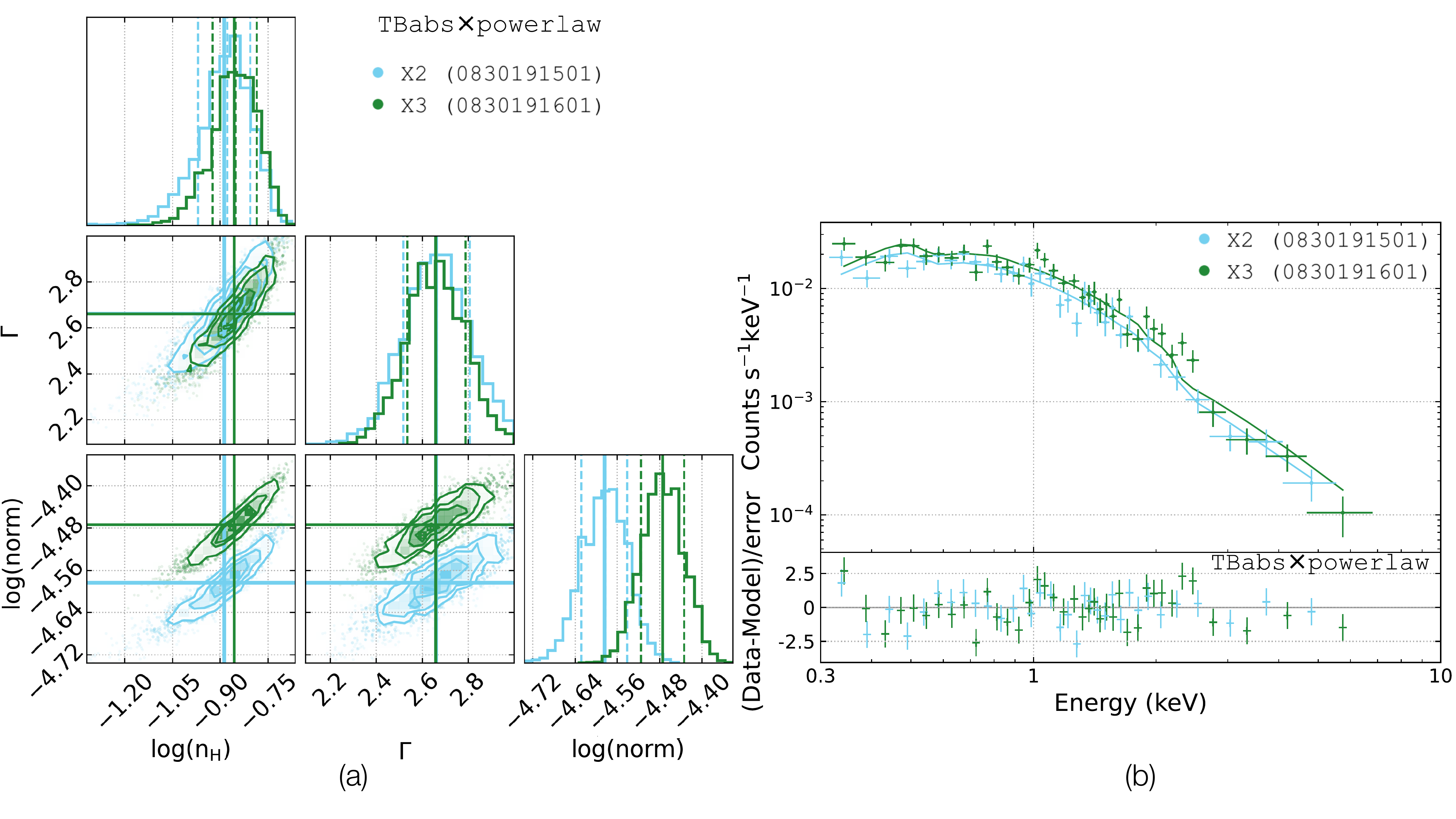}
        \caption{(a) Posterior distributions of the spectral parameters obtained with BXA for the absorbed power-law model, \texttt{TBabs}$\times$\texttt{powerlaw}, for X2 and X3. The one-dimensional histograms show the marginalized posterior distributions, while the contours show the joint posterior distributions between parameters. X2 and X3 are shown in blue and green, respectively. (b) Top: Background-subtracted XMM-Newton count spectra of X2 and X3 with the best-fitting absorbed power-law models shown as solid lines. Bottom: Fit residuals, defined as (data$-$model)/error. Both observations are only two days apart.}\label{fig:BXA_powerlaw}
    \end{figure*}

\subsection{X-ray Multi-Mirror Mission (XMM-Newton)}

We examined all publicly available XMM-Newton observations to search for any significant source detection. Among these, only the observations listed in \autoref{table:obs} resulted in detections and were consequently used for further analysis. The source was clearly detected in both X2 and X3. During these two observations, both EPIC-MOS1 and MOS2 operated in small window mode around the galactic center, leaving the source out of the field of view of the pointing. Only the EPIC-PN was operated in the full-frame mode. In the observation prior to both X2 and X3 (observation ID: 0830191401), the source location falls on the chip gap, making it impossible to assess the presence of any detection. 

We used the XMM-Newton Science Analysis System (\texttt{SAS} \citep{SAS04}) software (v.21.0.0) for data reduction following the online analysis threads. We used \texttt{epproc} to reprocess the observation data files from EPIC-PN and extract calibrated event files. Neither of the observations showed signs of a~flaring particle background. The default \texttt{RATE EXPRESSION} was used to extract the good time interval files and obtain filtered event files for further analysis. We extracted the background corrected light curves and spectra from $\sim$12$\arcsec$ centered around the source coordinates obtained from C2 with a~background region of $\sim$30$\arcsec$ selected from a~nearby source-free region. We followed the recommended selection expression and restricted the patterns only to single and double events (\texttt{PATTERN<=4}). We then used the \texttt{SAS} tools \texttt{rmfgen} and \texttt{arfgen} to generate redistribution matrices (rmf) and ancillary files (arf), respectively. We then grouped the obtained spectra with the \texttt{specgroup} task and binned to have at least 20 counts per bin. 

Additionally, we used the remaining publicly available XMM-Newton observations in which the source was not detected to obtain upper limits on the source flux before and after the first detection in X2. We first produced the sensitivity maps of each observation with the \texttt{SAS} tool \texttt{esensmap} and then used HEASARC's \texttt{webpimms}\footnote{\href{webpims}{https://heasarc.gsfc.nasa.gov/docs/software/tools/pimms.html}} tool \citep{Pimms} adopting the power-law model with $\Gamma =2.7$ and $N_{\rm H}=0.13 \times 10^{22}$~cm$^{-2}$(following the results of the spectral analysis from X2 and X3) to convert the corresponding values into flux in the 0.5-7~keV energy band. Prior to X2, the lowest flux value calculated from the source location was obtained from the 54.11 ksec observation (X1), where we estimated a~90 percent c.l. upper limit of its X-ray flux to be $F_{\rm x}= 1.79 \times {\rm 10^{-14}{~erg\ s^{-1} cm^{-2}}}$ with a~corresponding luminosity $ { L_{\rm x} = 1.57 \times 10^{38}} {~\rm{erg\ s^{-1}}}$. Following the last XMM-Newton detection (X3), the source decreased in the luminosity below the detection limit of the instrument and the upper limit on the flux and luminosity was even lower than prior to the detection with $F_{\rm x}= 1.53 \times {\rm 10^{-14}{~erg\ s^{-1} cm^{-2}}}$ with a~corresponding luminosity $ {L_{\rm x} = 1.34 \times 10^{38}} {~\rm{erg\ s^{-1}}}$.

We used OM data to search for a~possible optical/UV counterpart to M51 X*. We processed the XMM-Newton Optical Monitor (OM) data using the standard procedures in the Science Analysis System (SAS, v.21.0.0). We first ran the \texttt{omichain} task to process the OM images for source detection. We then ran the \texttt{omdetect} task with a detection threshold of 2$\sigma$ at the position of the X*. The final OM B-band image of M51 is shown in \autoref{fig:img}, which was then compared with the X-ray and UV images to search for possible optical counterparts.

\subsection{Neil Gehrels Swift Observatory (Swift)}

Swift-XRT has observed M51 extensively since 2005. We make use of this long-term monitoring to study the variability of M51 X* and constrain the duration and recurrence of its activity. We examined Swift-XRT observations spanning nearly three years before and after the first XMM-Newton detection, using images produced with the standard online tools provided by the UK Swift Science Data Centre \citep{Evans2009} Before the 2018 activity, the source showed no significant detections above the instrumental sensitivity. Shortly before the XMM-Newton observations, M51 X* was detected on 2018 June 12 (S1), although the low number of counts prevented the extraction of a spectrum suitable for spectral modeling.

The Swift-XRT observations obtained in August 2018 revealed a significantly brighter phase of X-ray activity, approximately two months after the XMM-Newton detections. The source was detected in two observations on August 14 and 18, yielding 4 and 7 counts, respectively. Given the low number of counts in the individual observations and their close temporal separation, we combined the two observations to improve the signal-to-noise ratio. The observed luminosity increased substantially compared with the June observations, indicating that the source entered the ultraluminous regime during this second episode. A subsequent observation on August 20 (S3) did not provide a significant detection and was used to derive an upper limit. The source was later detected by Chandra on August 31 (C2) at a considerably lower luminosity, indicating a rapid decline following the second episode. Because the individual Swift-XRT observations contain only a small number of source counts, we did not attempt detailed spectral modeling of these data. Instead, the Swift observations were used primarily to characterize the long-term evolution of M51 X* and to constrain the timing of the episode.

The Ultra-violet Optical Telescope (UVOT) onboard Swift did not detect any significant activity associated with the source position.

\subsection{Hubble Space Telescope (HST)}

Unfortunately, HST did not observe M51 during or shortly after the outburst. In order to investigate the region, we used HST observations from September 2014 with the WFC3/UVIS F275W filter (PropID: 13364, PI Calzetti). We selected a~region of 2\arcsec centered around the source coordinates obtained by the Chandra detection and assumed an error margin of 0.5\arcsec. We then performed a~PSF photometry with the DAOPHOT package on \texttt{IRAF} with a~2\,pixel ($0\farcs8$) aperture. The bottom right image in \autoref{fig:img} shows the population of potential point sources we were able to identify within this region, highlighted with smaller yellow circles.

\subsection{Astrorivelatore Gamma ad Immagini LEggero \\ (AGILE)}


We analyzed the AGILE-GRID data between June 12 and August 31, 2018, during which the M51 region was continuously covered in sky-scanning mode. We used \texttt{agilepy} (v.1.6.5)\footnote{https://agilepy.readthedocs.io/en/1.6.5/} with the AGILE-GRID Build 26 package, FM3.119 calibrated filter, H0025 response matrices, and the consolidated ASDCSTDk archive. Standard data selections were applied, including South Atlantic Anomaly removal, an $80^\circ$ Earth-albedo cut, and an off-axis angle $<60^\circ$. To account for the background, we included all known sources from the second AGILE catalog \citep[2AGL][]{bulgarelli_2019} within $20^\circ$ of M51 X*. No significant gamma-ray source was detected at a position consistent with M51 X*.

\section{Data Analysis and Results}\label{results}

\subsection{Long Term Evolution}

We analyzed publicly available XMM-Newton, Chandra, and Swift observations to constrain the duration and recurrence of the source activity. Swift monitored M51 extensively between 2005 and 2025, providing long-term coverage before and after the 2018 detections. \citet{Brightman20} analyzed a $\sim$1.5-year Swift campaign beginning in May 2018 but did not detect M51 X*, likely because observations were stacked over intervals longer than the source activity. In contrast, Swift data restricted to the active periods confirm the source detection, although the short exposures limit further analysis.

The long-term light curve in \autoref{fig:lcurve} suggests possibly a continuous episode of X-ray activity in 2018. It began with the Swift-XRT detection S1 on June 12, followed by XMM-Newton observations X2 and X3 on June 13 and 15, reaching $7.8\times10^{38}\,\mathrm{erg\,s^{-1}}$. The peak of the outburst was detected by Swift-XRT on August 14-18 (S2), with S2 reaching the ultraluminous regime. The source entered a decline on August 20 (S3) and had faded to $8\times10^{37}\,\mathrm{erg\,s^{-1}}$ by the Chandra observation on August 31 (C2). It is worth noting that this activity could also be identified as a double-peak event due to the lack of Swift-XRT detections between X3 and S2 but short snapshots obtained during this period limit the sensitivity of the instrument to lower flux values than S2.

\subsection{Spectral Analysis}

We performed the spectral analysis with the \texttt{HEASOFT} (v.6.31.1) tool \texttt{XSPEC}  \citep[v.12.13.0c, ][]{Arnaud96}. We utilized the Cash statistic, namely C-stat \citep{Cash79}, to account for the Poisson distribution of the observed photons in such low count rate spectra. We performed the initial analysis with \texttt{XSPEC} to explore possible contributions to the model and make a~preliminary model comparison. To improve the parameter estimation and perform a~more comprehensive model comparison, we made use of the Bayesian Analysis Package \citep[BXA, ][]{Buchner14}, which is a~Bayesian framework utilizing the nested sampling algorithm UltraNest \citep{Buchner16,Buchner19,Buchner21} to explore the parameter space of any given model and provided priors. We defined log-uniform priors for $\rm N_{H}$ and normalization parameters of both models, and a~uniform prior for $\Gamma$  with the following parameter ranges for XMM-Newton observations X2 and X3: $\log(N_{\rm H})=[-1.3,-0.7]$, $\Gamma=[2.0,3.0]$, and $\log(\mathrm{norm}_{\rm powerlaw})=[-4.8,-4.4]$. For fits including the \texttt{diskbb} component, we adopted $kT_{\rm in}=[0.1,1.2]$~keV and $\log(\mathrm{norm}_{\rm diskbb})=[-2.4,1.2]$. For the Chandra C2 spectrum, we similarly adopted a uniform prior for the photon index, $\Gamma=[1.0,5.0]$, and a log-uniform prior for the power-law normalization, $\log(\mathrm{norm}_{\rm powerlaw})=[-5.8,-4.8]$, while $N_{\rm H}$ was fixed at $0.13\times10^{22}\,\mathrm{cm^{-2}}$. The XMM-Newton priors were guided by preliminary XSPEC fits, while the broader Chandra C2 photon-index prior encompassed physically plausible values. However, the limited photon statistics and prior-dependent uncertainties prevent us from establishing significant spectral evolution relative to the XMM-Newton observations.

\begin{figure*}[h!]
\centering
\hspace{-0.5cm}
        \includegraphics[width=\textwidth]{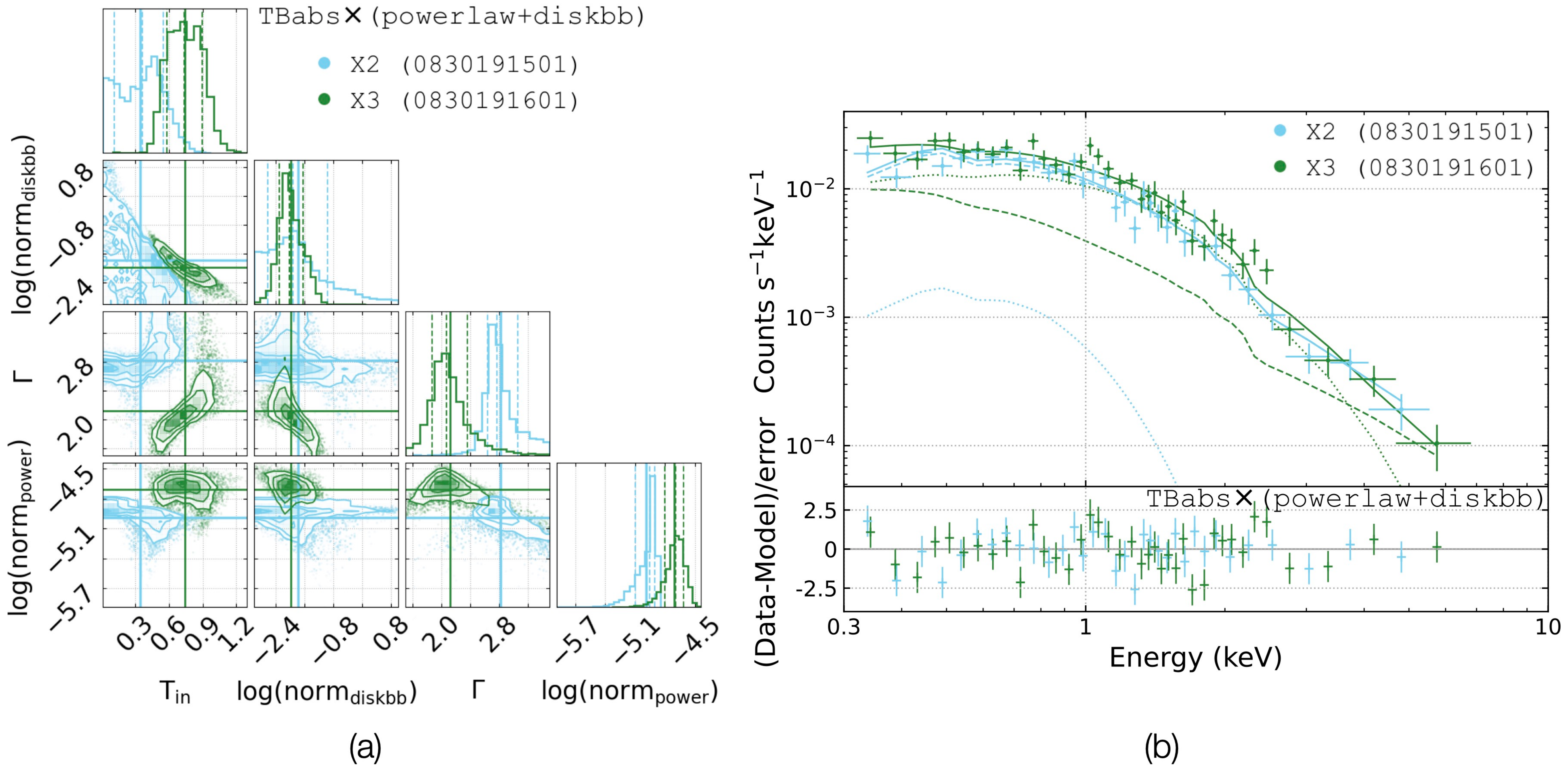}
        \caption{(a) Posterior distributions of the spectral parameters obtained with BXA for the \texttt{TBabs}$\times$(\texttt{powerlaw}+\texttt{diskbb}) model for X2 and X3, only two days apart. The one-dimensional histograms show the marginalized posterior distributions, while the contours show the joint posterior distributions between parameters. X2 and X3 are shown in blue and green, respectively. (b) Top: Background-subtracted XMM-Newton count spectra of X2 and X3 with the best-fitting models shown as solid lines; the individual \texttt{powerlaw} and \texttt{diskbb} components are shown as dashed and dotted lines, respectively. Bottom: Fit residuals, defined as (data$-$model)/error.}\label{fig:BXA_diskbb}
    \end{figure*}

\begin{figure*}[h!]
\centering
\hspace{-0.5cm}
        \includegraphics[width=\textwidth]{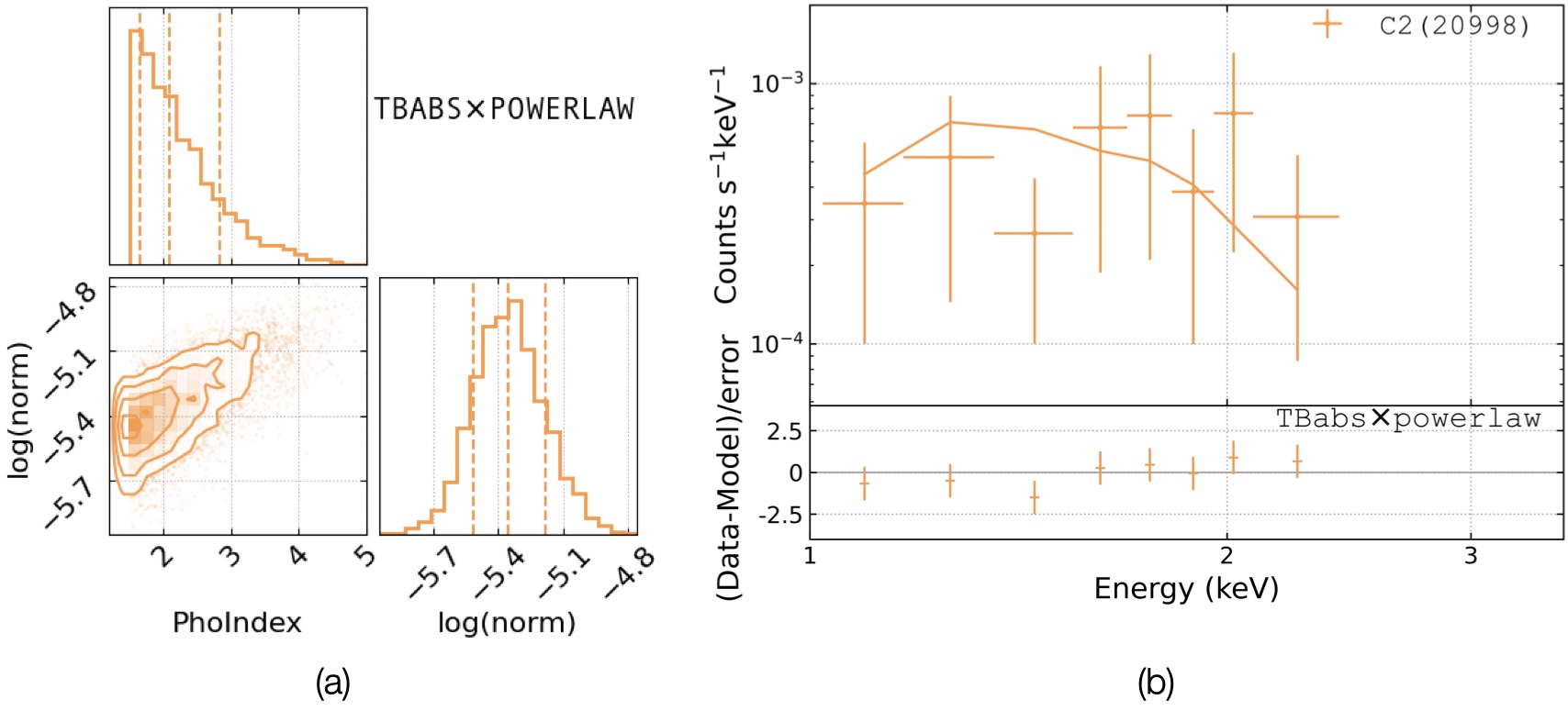}
        \caption{(a) Posterior distributions of the spectral parameters obtained with BXA for the Chandra C2 spectrum using an absorbed power-law model, \texttt{TBabs}$\times$\texttt{powerlaw}. The one-dimensional histograms show the marginalized posterior distributions, while the contours show the joint posterior distributions between parameters. (b) Top: Background-subtracted Chandra count spectrum of C2 with the best-fitting absorbed power-law model shown as a solid line. Bottom: Fit residuals, defined as (data$-$model)/error.}\label{fig:BXA_powerlaw_chan}
    \end{figure*}
    
First, we applied a~simple absorbed power-law model (\texttt{TBabs$\times$powerlaw}) to both spectra (X2 and X3) using the default element abundances and cross-sections \citep{anders1989,verner1996} with all parameters left free to vary. While the spectral shapes remained similar with only a~slight increase in the flux, observation in X3 produced worse fit test statistics with $\chi ^{2}$/d.o.f=57.83/36. The left panel of \autoref{fig:BXA_powerlaw} shows the corner plot of distributions of parameters in \texttt{TBabs$\times$powerlaw}. ${\rm{{\chi}^2/d.o.f.}}$ values of both fits are presented in \autoref{tab:results}.

Given the increase in the $\chi ^{2}$/d.o.f. with increasing flux, we attempted an additional fit with a~multicolor accretion disk model in \texttt{TBabs$\times$(powerlaw+diskbb)}. While the fit test statistics improved, the residuals in lower energies did not show significant differences between the two observations. The left panel of \autoref{fig:BXA_diskbb} shows the corner plot of distributions of parameters in \texttt{TBabs$\times$(powerlaw+diskbb)}, and we provide the results of both fits in \autoref{table:spec}. An F-test comparison between the two models yields $F=1.23$ with a probability of $p=0.28$ for X2, indicating that the additional disk component is not statistically required. For X3, on the other hand, the test returns $F=9.46$ with $p=0.004$, suggesting a statistically significant improvement when including the \texttt{diskbb} component.

In addition to the phenomenological models \texttt{powerlaw} and \texttt{powerlaw+diskbb}, we tested different Comptonisation models (\texttt{simpl}~\citep{Steiner09}, \texttt{Nthcomp}~\citep{Zdziarski96,Zycki99}, \texttt{compTT}~\citep{Titarchuk94} in the \texttt{xspec} environment) to provide a~more physically motivated modeling of the high-energy contribution to the continuum. However, given the soft nature of both spectra, we found no significant improvement to the fit. Motivated by previous works \citep{Stobbart06, Gladstone09,Koliopanos17, Walton21}, we also tested the addition of a~two-thermal-components model by including two disk components (\texttt{TBabs$\times$(powerlaw+diskbb+diskbb)} and \texttt{TBabs$\times$(diskbb+diskbb)} as total models), which are usually characterized by two nodes of temperatures around 0.3 keV and 3 keV.  The latter is used to describe an inner accretion flow, and the low-temperature component is assumed to arise from either the outer regions of the accretion disk or the photosphere of a~strong wind \citep{King03}. While this approach is preferred for super-Eddington accretion disks, the addition of another disk component provided only a~statistically insignificant decrease in the fit test statistics, which can be attributed to the addition of free parameters to the model. Both disk temperatures remained relatively low compared to accretion disk temperatures observed in black hole XRBs, with the highest temperature component reaching only $\mathrm{k}T_{\rm in} \sim$ 0.7 keV in X3. 

We also fitted the Chandra C2 spectrum with an absorbed power-law model. Due to the low number of counts, we fixed the absorption at $N_{\rm H}=0.13\times10^{22}$ cm$^{-2}$. The fit yielded $\Gamma=2.09^{+0.74}_{-0.43}$, consistent within the uncertainties with the soft spectra observed with XMM-Newton. The left panel of \autoref{fig:BXA_powerlaw_chan} shows the corner plot of distributions of parameters in \texttt{TBabs$\times$powerlaw}, and we provide the results of the fit in \autoref{table:spec}.

For the Swift-XRT observations, fluxes were estimated assuming an absorbed power-law model with $N_{\rm H}=0.13\times10^{22}\,\mathrm{cm^{-2}}$ and $\Gamma=2.7$, based on the XMM-Newton spectral fits. As the lower uncertainty on the S3 flux was consistent with zero, we conservatively report this observation as an upper limit. We note that the photon index may have evolved between the XMM-Newton and Swift-XRT observations, introducing additional uncertainty in the inferred luminosity. However, given the estimated luminosity, moderate variations in the photon index are unlikely to affect the classification of M51 X* as a ULX candidate.

\subsection{Multiwavelength Counterparts}

\subsubsection{Optical, ultraviolet and infrared}
We extended our analysis to search for possible multi-wavelength counterparts within the region corresponding to the Chandra detection. XMM-Newton-OM provided no detection during the outburst corresponding to an estimated $B$-band upper limit of $B \gtrsim 21$ mag, although it is most probably a~result of the instrument's limited sensitivity to detect a~companion at such large distances. 

Even in the case of a~possible detection, it is unlikely that the contribution will be observed only from the companion; however, it could correspond to a~populated region, which can be better resolved with the HST. Unfortunately, HST did not observe the region where M51 X* was detected on the same temporal frame as Chandra and XMM-Newton detections, either during or shortly after the outburst. We could not identify any strong candidates for counterparts in the optical range, although in different temporal frames, several sources were located within the Chandra aperture of $2$\arcsec{} in the UV (see \autoref{fig:img}). When the metallicity of the galaxy is considered \citep{Wei2020} in combination with the distance, we were able to identify a~total of 10 possible counterparts compatible with the OB-type stars and magnitudes between 23 and 26\,mag. However, we cannot draw any conclusions about these possible counterparts due to missing exposures in other filters, which makes it impossible to create a~color-magnitude diagram. 

\subsubsection{Gamma Rays}

No high-energy counterparts were identified with an upper limit $F = 8.48 \times 10^{-11}\,\mathrm{erg\,s^{-1}\,cm^{-2}}$ in the 100 MeV--50 GeV range with AGILE. Although this may be partly due to instrumental sensitivity, the emission of the outburst is dominantly constrained to X-rays.

\begin{table*}[h!]
\centering
\renewcommand{\arraystretch}{1.4}
\begin{threeparttable}
\resizebox{\textwidth}{!}{%
\begin{tabular}{cccccccc}
\hline
DataID & Model\tnote{a} & ${\rm{\chi^2/d.o.f.}}$ & ${\rm N_H}$ & $\Gamma$ & $\mathrm{k}T_{\rm in}$ & Flux\tnote{b} & ${\rm L_x}\tnote{c}$ \\
& & & $\left(10^{22}\,\mathrm{cm}^{-2}\right)$ & & (keV) & $\left(10^{-14}\,\mathrm{erg\,s^{-1}\,cm^{-2}}\right)$ & $\left(10^{38}\,\mathrm{erg\,s^{-1}}\right)$ \\
\hline
X2 & \texttt{TBabs$\times$powerlaw} & 34.66/31 & $0.13^{+0.03}_{-0.02}$ & $2.68^{+0.14}_{-0.15}$ & -- & $7.99^{+0.35}_{-0.33}$ & $7.04^{+0.31}_{-0.29}$ \\
& \texttt{TBabs$\times$(powerlaw+diskbb)} & 33.29/30 & 0.13 (fixed) & $2.77^{+0.28}_{-0.14}$ & $0.36^{+0.18}_{-0.25}$ & $7.97^{+0.58}_{-0.60}$ & $7.02^{+0.51}_{-0.52}$ \\
\hline
X3 & \texttt{TBabs$\times$powerlaw} & 57.83/36 & $0.14\pm0.02$ & $2.67^{+0.13}_{-0.12}$ & -- & $8.62^{+0.51}_{-0.53}$ & $7.76^{+0.45}_{-0.46}$ \\
& \texttt{TBabs$\times$(powerlaw+diskbb)} & 43.53/35 & 0.13 (fixed) & $2.07^{+0.29}_{-0.19}$ & $0.73^{+0.16}_{-0.15}$ & $8.61^{+0.50}_{-0.52}$ & $7.58^{+0.44}_{-0.45}$ \\
\hline
C2 & TBabs$\times$powerlaw & 4.66/6 & 0.13 (fixed) & $2.09^{+0.74}_{-0.43}$ & -- & $0.95^{+0.38}_{-0.51}$ & $0.80^{+0.30}_{-0.40}$ \\
\hline
\end{tabular}%
}
\caption{\label{table:spec} Results of the spectral fits with BXA used in this paper. Here ${\rm N_H}$ is the absorption column density, $\mathrm{k}T_{\rm in}$ is the temperature of the inner disk, and $\Gamma$ is the power-law photon index.}
\label{tab:results}
\begin{tablenotes}
\item[a] Both fits with XSPEC and BXA were performed with C-stat fit statistics, and Chi-squared fit test statistics are presented here to evaluate the goodness of fit between the two observations.
\item[b] Observed flux in the 0.5--10.0 keV energy range calculated using the convolution model \texttt{cflux} within XSPEC.
\item[c] Observed luminosity in the 0.5--10.0 keV energy range calculated using the flux values obtained from spectral fitting, assuming a distance of 8.58 Mpc.
\end{tablenotes}
\end{threeparttable}
\end{table*}

\subsection{Timing Analysis}

We carried out a~detailed timing analysis to look for coherent pulsations in the \textsl{XMM-Newton} EPIC-pn data. We did not use data from other instruments or satellites as they do not provide sufficient time resolution for pulsation search with periods $P$ around 1\,s, which is the period typically found for ULX pulsars.
First, we performed a~Fast Fourier Transform analysis on the light curves with a~time resolution of 74.3\,ms (the frame time of the EPIC-pn detector) between $10^{-4}$-$10^{-7}$\,Hz. We did not find any significant peaks in either the X2 or the X3 data.
We then employed an accelerated pulse period search by performing epoch-folding over a~grid in period $P$ and period derivative $\dot P$ space between $P=0.25$-$1.25$\,s and $\dot P=-1.5\times10^{-9}$-$1.5\times10^{-9}\,\text{s}^{-1}$. These ranges cover the values typically found in PULXs \citep[e.g., ][]{isr+17a, fur+16}. Again, no significant pulsations could be identified in either X2 or X3 data.

To place upper limits on the pulsed fraction (PF), we simulated event files based on an input event list with the same average count rate as the real data. We then added sinusoidal pulsations to these simulated files, resulting in a~pulsed fraction $\text{PF}_\text{sim}$. The simulations follow the method described in \citet{Furst21} and are based on the method used in the \texttt{Stingray} software package \citep{huppenkothen2019}.
Specifically, we simulated event lists with pulsed fractions for each observation in the range $0.1 \leq \text{PF}_\text{sim} \leq 0.9$ with steps of 0.01. For each value of $\text{PF}_\text{sim}$ we generated 100 statistically independent event lists and searched within each list for pulsations using the epoch folding technique. We did not include $\dot P$ in these simulations as it does not influence the obtained upper limit. We defined the upper limit on the PF as the value where at least 90\% of all simulated event lists provide detection of the pulse with at least 99.5\% significance.  We found an upper limit of $\text{PF}<32$\% for X2 and  $\text{PF}<29$\% for X3.

\section{Discussion}\label{discussion}

\subsection{Spectral Properties and Variability}

Except for the slight increase in the X-ray flux, no significant variability was observed in either the X2 or X3 spectra. This change resulted only in a~small difference (see the corner plots in \autoref{fig:BXA_powerlaw} and \autoref{fig:BXA_diskbb}) in the normalization parameter of \texttt{powerlaw} when fitted with a~simple absorbed power law (\texttt{TBabs$\times$powerlaw}, \autoref{fig:BXA_powerlaw}). Both spectra remained soft throughout the increase in the flux with $\Gamma \sim \rm 2.6$. 


\subsection{Timing Variability and Geometry of the System}

The timing properties of M51 X* provide limited constraints on the geometry and nature of the accretor. The X2 and X3 light curves show no significant short-term variability, eclipses, or pulsations. The absence of eclipses could indicate a relatively low inclination, but the limited temporal coverage prevents such a conclusion if the orbital period is longer than the individual observations. Similarly, the lack of continuous coverage prevents us from constraining the orbital period.

\paragraph{Neutron-star accretor and pulsations.}
No significant pulsations were detected in either X2 or X3. This does not exclude a neutron-star accretor, since X-ray pulsations are suppressed in at least 60\% of ULXs known to host neutron stars \citep{Misra24}. Although neutron stars may constitute a substantial fraction of the ULX population \citep{King16}, their pulsations can be difficult to detect depending on the accretion geometry and viewing angle. The upper limits on the pulsed fraction of M51 X* are comparable to the pulsed fractions observed in some PULXs. For example, P13 in NGC 7793 exhibits pulsed fractions comparable to or higher than our upper limits, demonstrating that pulsations can be detectable under favorable geometric conditions \citep{Furst21}. Thus, the non-detection of pulsations in M51 X* does not by itself distinguish between a neutron-star and black-hole accretor.

\paragraph{Super-Eddington accretion and geometrical beaming.}
At accretion rates close to or above the Eddington limit, the inner accretion flow can become geometrically thick and form a radiation-pressure-dominated funnel \citep{1988ApJ...332..646A}. Radiation emerging through this funnel can be geometrically beamed, increasing the apparent luminosity along favorable lines of sight \citep{King20,Abarca21,2025ApJ...982...95K}. For M51 X*, such beaming could contribute to the high apparent luminosity reached during the second episode of activity. However, a nearly face-on view of the inner accretion flow would generally expose hotter inner regions, whereas M51 X* shows a relatively soft X-ray spectrum. The available data therefore provide no clear evidence that strong geometrical beaming dominates the observed luminosity.

At higher inclinations, super-Eddington disk winds can obscure the hotter inner regions and produce softer observed spectra \citep{Pinto2023}. The soft spectrum of M51 X* could therefore be consistent with viewing a super-Eddington flow at moderate-to-high inclination. However, if the lack of pulsations were caused primarily by strong obscuration along the line of sight, a larger absorbing column might be expected than inferred from the XMM-Newton spectra. The current observations consequently do not provide a unique constraint on the inclination or degree of geometrical beaming.

\paragraph{Black-hole accretor.}
A black-hole accretor provides an alternative explanation for the absence of pulsations. For a stellar-mass black hole, the observed luminosity could be produced by accretion near or above the Eddington limit, potentially accompanied by a geometrically thick inner disk and outflows. An intermediate-mass black hole would instead allow the observed luminosity to remain sub-Eddington and would generally shift the characteristic disk emission toward lower energies. However, the current spectral and multiwavelength constraints do not allow us to distinguish between these possibilities or provide evidence specifically favoring an intermediate-mass black hole.

\paragraph{Transient behavior and inclination.}
The rapid transient behavior provides an additional constraint on these scenarios. One possibility is that changes in visibility are caused by a high-inclination or eclipsing configuration. However, the absence of eclipses during X2 and X3 does not strongly constrain this scenario because the orbital period is unknown and the observations provide only sparse coverage. More importantly, the large luminosity changes observed between the two episodes of activity and the intervening non-detections cannot presently be attributed uniquely to geometrical effects. Changes in the intrinsic accretion rate, obscuration, or viewing geometry could all contribute to the observed variability.

Overall, neither the timing nor spectral properties uniquely identify the nature or geometry of M51 X*. The absence of pulsations does not exclude a neutron-star accretor, while the soft spectrum and high luminosity can be accommodated by different viewing geometries in super-Eddington accretion models. A black-hole accretor also remains viable. The sparse temporal coverage, particularly between the two episodes of activity, prevents us from determining whether the rapid variability is primarily driven by changes in accretion rate or geometry. Future observations during an active episode will be required to distinguish between these scenarios.

\subsection{Nature and Mass of the Accretor}

High luminosities of observed ULXs raised the assumption that these objects should be intermediate-mass black holes. At the moment, there is growing support for a~theory of accretion onto neutron stars powering these sources \citep{Koliopanos17}. However, this simply does not provide enough grounds to rule out the black hole nature of all ULXs easily. It might not be appropriate to rely solely on the lack of pulsation observations or the propeller regime-like variability to infer the existence of a~black hole accretor. In this section, we consider potential scenarios to investigate the nature of the compact object and the structure of the accretion disk in M51 X* based on available observed properties.

\subsubsection{Stellar-mass Compact Object: Black Hole vs. Neutron Star}


\textit{A stellar-mass black hole}, typically in an XRB system which undergoes an outburst each observed spectral state is characterized by a~certain timescale \citep{Dunn10} that is about several months. Outbursts on much shorter timescales than the case of M51 X* are referred to as "failed" as the source fails to transition from the low/hard state (LHS) to the high/soft state (HSS). However, these failed outbursts are characterized by a~short-lived increase in flux accompanied by much harder spectra ($\Gamma \sim 1.7$). In the case of a~typical black hole XRB, the transition from a~quiescent to an ultraluminous state, following an increase in the observed flux by several orders of magnitude, is not expected to take place on timescales similar to the outburst observed in M51 X*. Additionally, the transition from an LHS to HSS observed in many black hole XRBs is expected to have signatures of accretion state changes characterized by the hardness of the spectrum, while M51 X* showed no signs of spectral variability throughout the outburst, with a~somewhat soft spectrum ($\Gamma \sim \rm 2.6$) accompanied by a~slight increase in the observed flux. If these two observations are to be taken as the \say{awakening} of a~black hole accretor, then the rise in the X-ray flux should have been accompanied by a~hard spectrum with $\rm \Gamma \sim 1.7$ which stays constant before entering the intermediate state \citep[see ][ for a~more detailed discussion]{Dunn10,Dunn11}. On the other hand, sudden softening of the spectrum throughout the rise from the LHS is not that unexpected in the case of a~Compton-thick disk wind, as observed in GRO J1655-40 \citep{Uttley15}. This scenario is hard to either prove or rule out because of the significantly reduced S/N ratio at higher energies, where the signatures of absorption due to a~disk wind are expected. The anomalous case of GRO J1655-40 is so far observed to be isolated, and the accretion rate through that shift remained well below the Eddington limit for a~stellar-mass black hole accretor \citep{Dunn10,Dunn11,Uttley15,Yilmaz23}. 

The sudden increase in the flux, accompanied by the soft nature of the spectra observed in X2 and X3, could be interpreted as the presence of an accretion disk. However, currently available accretion disk models are not able to explain the short-lived outbursts observed in several ULXs \citep{Fabrika15}. For a~stellar-mass black hole accretor powering M51 X*, the accretion disk would be geometrically very thick due to the high mass accretion rate and emit a~rather soft spectrum. The presence of a~geometrically and optically thick disk around a~black hole may lead to strong beaming while keeping the spectra rather soft \citep{Wielgus22}.

Another possible scenario for transient ULXs with a~superorbital period is the presence of a~precessing warped disk that is inclined with respect to the rotational axis of the central object. The Lense-Thirring precession would lead to a~periodical obscuration of the central regions, as was observed in similar transient NGC 925 ULX-3 \citep{2020ApJ...891..153E}. However, in the case of M51 X*, we do not observe any periodicity, which could be the result of the lack of long-term observing campaigns of M51. The sharp decay and a~single occurrence of the source are strongly in contradiction with this scenario, and we can almost certainly rule it out. 

\textit{Neutron star} powered ULXs, on the other hand, are found to be more variable, and neutron stars with stronger magnetic fields are observed to be short-lived \citep{Thompson95}. Considering M51 is a~star-forming galaxy and the association of neutron stars with recent star-formation history (e.g. \citet{Heger03} and \citet{Cerda-duran18} for a~more recent review), it is not straightforward to rule out the neutron star nature of the accretor. Invoking a~neutron star as the accretor could ideally provide a~plausible scenario in which the accretion stops abruptly. A~neutron star will only be able to accrete matter when the corotation radius 
is located outside the magnetospheric radius. 
In this case, if there is a~decrease in the accretion rate, the magnetospheric radius can become larger and eventually enter the propeller regime and shred the accretion disk apart \citep{White88, cui97}. Another possibility is that continued accretion transfers angular momentum to the neutron star, gradually reducing its spin period and causing the corotation radius to move inward \citep{Ghosh79}. However, the propeller regime is expected to be dominantly driven by the changes in the accretion rates instead of the changes in the spin period of the neutron star, while the observed spectral and timing properties of such systems point towards a~more complicated picture \citep{Middleton23}.

The propeller effect, however, is usually used to explain the bimodal distribution of the observed flux in some ULXs over a~longer period of time, spanning months or even years. Although M51 X* was observed through only two episodes of activity, we cannot definitively consider these an isolated event and rule out the bimodal distribution due to the lack of more frequent long-exposure observations. While strongly magnetized neutron stars can maintain efficient super-Eddington accretion, weakly magnetized neutron stars are argued to be able to reach super-Eddington regimes, launching powerful outflows instead \citep{King16}. At the same time, simulations of \citet{Abarca21} and \citet{2025ApJ...982...95K} showed that the beamed emission of an accreting neutron star could reproduce the apparent luminosity of $\sim 100 \,L_{\mathrm{Edd}}$ even for slightly super-Eddington accretion rates and not strongly magnetized neutron stars.

Another explanation for the observed properties may be that the source is a~neutron star X-ray binary harboring a~Be star as a~companion \citep[BeXRB, see ][for a~more detailed review and the references therein]{Reig11}. BeXRBs undergo two types of outbursts: (i) Type-I, periodic or quasi-periodic short outbursts spanning a~small fraction of the orbital period where the peak X-ray luminosity is usually below $\sim \rm 10^{37}~erg\ s^{-1}$; (ii) Type-II, relatively longer (lasting longer than 100 days) and less frequent outbursts that show a~significantly larger increase by three to four orders of magnitude in the X-ray luminosity and reaching the Eddington limit for a~neutron star accretor. 
The outburst observed in M51 X*, however, appears relatively short in duration, but again, given the sparsity of the data, we cannot exclude this scenario.

\subsubsection{Background Tidal Disruption Event or Active Galactic Nucleus}

The absence of bright optical or UV counterparts with flux densities comparable to the observed X-ray emissions makes it unlikely that the source is a~background quasar or blazar \citep{shang11}.
On the other hand, isolated bursts on the timescales of 2-3 months spanning nearly two decades of observing time of their host galaxies can indicate the rise and decline of a~tidal disruption event (TDE).  

A TDE occurs when a~star approaches close enough to a~massive black hole whose tidal field overcomes the self-gravity of the star and tears it apart which generates a~bright signal with a~characteristic power-law decay that can last from days to years \citep{hills1975,rees1988,alexander2005}. TDEs have been detected across the whole electromagnetic spectrum but mostly in optical, UV, and X-ray bands \citep[see][for a~review]{gezari21}. 

\begin{figure}
        \centering
        \includegraphics[width=0.5\textwidth]{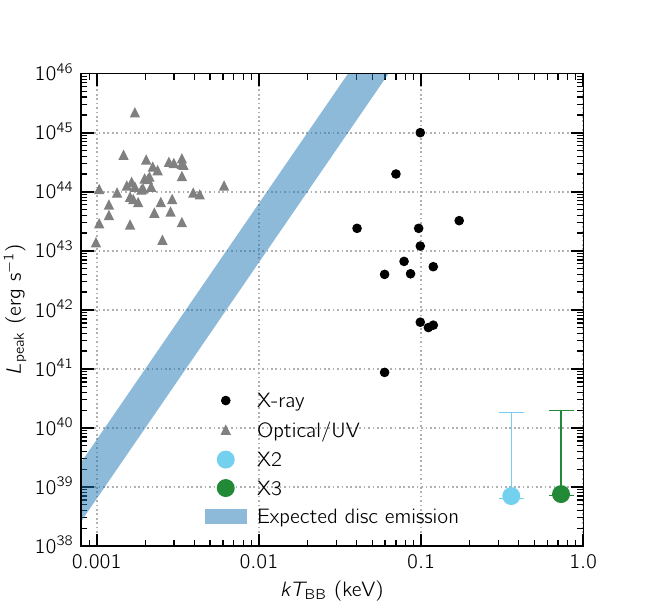}
        \caption{
            Peak luminosity as a~function of measured blackbody temperature for 53 tidal disruption events compiled by \protect\cite{gezari21}. We indicate the peak luminosities obtained from X2 and X3 in blue and green, respectively. We emphasize that the error bars shown in the figure are to illustrate a~possible range of luminosities if the source were a~background TDE, calculated based on the observed redshifts of the population of observed TDEs \citep{arcavi2014,french2016}.
            Grey triangles and black dots represent events detected in optical and X-ray bands, respectively. The diagonal blue-shaded region corresponds to the expected luminosity irradiated by an accretion disk of a~Solar-like star disrupted by a~black hole of mass in the range $10^0-10^8~\text{M}_{\odot}$ with accretion efficiency $\eta\sim0.01$-$0.1$.}
        \label{fig:tde}
    \end{figure}

We use the sample collected by \cite{gezari21}, who compared the peak luminosity with the blackbody temperature of TDEs \citep[see Figure 2 in][]{gezari21}, to compare our source to it. Even when assuming a~distance to the source at 8.58 Mpc (as in M51), the X-ray luminosity of M51 X* is about $7.8\times10^{38}~\text{erg}~\text{s}^{-1}$, which is significantly smaller than the X-ray observed sample ($10^{41}-10^{45}~\text{erg}~\text{s}^{-1}$). Given that the majority of the observed population of TDEs is observed in post-starburst galaxies at the relatively lower end of the redshift ($z\lesssim0.5$) scale \citep{arcavi2014,french2016}, we can position both X2 and X3 in the region of low peak luminosities ($10^{38}-10^{40}~\text{erg}~\text{s}^{-1}$). Following our results, these observations are significantly lower than the typical luminosity of X-ray-detected TDEs (about two orders of magnitude), see \autoref{fig:tde}.

On the other hand, the rapid variability of M51 X* alone cannot exclude a background active galactic nucleus (AGN), as AGN can show variability over a wide range of timescales. In particular, \say{changing-look AGN} can vary on timescales from hours to years, either through intrinsic changes in the accretion flow or variations in the line-of-sight obscuration \citep{Matt03, LaMassa15, Graham20, Ricci22}. In the case of M51 X*, the lack of strong variations in the measured absorbing column disfavors a scenario in which the observed variability is driven primarily by changing obscuration. However, intrinsic AGN variability cannot be excluded based on the X-ray variability alone. Furthermore, the location of M51 X* within a spiral arm of M51 supports its association with the host galaxy and identification as a ULX. Together with the lack of a bright optical or UV counterpart at the position of M51 X*, these properties make a background AGN interpretation less likely, although they do not rule it out. 

\section{Conclusions}\label{conclusions}

In this paper, we report the discovery and analysis of a~new transient X-ray source M51, M51 X*, based on archival X-ray observations from XMM-Newton, Chandra, and Swift-XRT. Our results can be summarized as follows:

\begin{itemize}
    \item M51 X exhibited strong X-ray variability during its 2018 period of activity. Initially detected in June by Swift-XRT and XMM-Newton, the source reached a luminosity of $7.8\times10^{38}\,\mathrm{erg\,s^{-1}}$ in the XMM-Newton observations. Subsequent Swift-XRT observations in August revealed a significantly brighter phase, with a luminosity reaching approximately $(5.2\times10^{40}\,\mathrm{erg\,s^{-1}}$, firmly placing the source in the ultraluminous regime. The source subsequently faded to approximately $8\times10^{37}\,\mathrm{erg\,s^{-1}}$ in the Chandra observation on August 31 (C2). The sparse temporal coverage prevents us from determining whether the observed activity represents a single outburst or multiple episodes.
    
   \item The August Swift-XRT observations establish M51 X* as a transient ULX, with an observed luminosity exceeding the canonical ULX threshold of $10^{39}\,\mathrm{erg\,s^{-1}}$ \citep{Kaaret17, Fabrika21}. The large luminosity changes between detections and non-detections demonstrate the strongly transient nature of the source, although the sparse sampling prevents us from determining whether the two episodes arise from the same underlying outburst cycle.

    \item The source lies 5.27\arcsec{}  from a~star cluster in M51, which is consistent with the location of other ULXs found in starburst galaxies. No optical or UV counterparts were detected in simultaneous XMM-Newton and Swift UVOT observations, ruling out a~possible background AGN or foreground star scenario.

    \item No significant variability in spectral shape was observed during the outburst as the photon index remained consistently soft ($\Gamma \sim 2.6$). Timing analysis revealed no pulsations in the brightest epochs with upper limits to the pulsed fraction of PF $<32$\% (X2) and $<29$\% (X3). Furthermore, no short-term variability was detected in the light curves.

    \item No high-energy counterparts were identified with an upper limit $ {F = 8.48 \times 10^{-11}} {~\rm{erg\ s^{-1} cm^{-2}}}$ in the 100 MeV - 50 GeV range with AGILE. Although this may be partly due to instrumental sensitivity, the emission of the outburst is dominantly constrained to X-rays.

    \item The absence of pulsations, with upper limits on the PF near the PULX regime, may offer some constraints on the system geometry, but does not clarify the underlying mechanism driving the transient behavior. There is no evidence to support the scenario in which a~geometrically thin disk can persist under super-Eddington conditions without significant geometrical beaming.

\end{itemize}

The new highly luminous X-ray transient M51 X* highlights a~growing gap in our understanding of critically accreting systems. The observed variability suggests that magnetic instabilities may disrupt the inner accretion flow, possibly halting accretion. We suggest that magnetic processes, such as those in the magnetically arrested disk, \citep[MAD,][]{2003PASJ...55L..69N} or propeller regimes, could partly explain the behavior, although further theoretical and numerical studies are needed. However, limited long-term monitoring of M51 restricts our ability to determine the origin of the outburst disruption. Future observation campaigns focused on the newly discovered source may lead to uncovering the mechanism of the outburst as well as help to better establish the source properties and nature.\\

\begin{acknowledgements}
     The authors thank Luigi Stella, Ji\v{r}\'{i} Hor\'{a}k, and K. Yavuz Eksi for the insightful discussions. A.Y. and J.S. acknowledge GA\v{C}R project No. 21-06825X for the support. D.L. also acknowledges the GA\v{C}R grant no. 25-16928O and the internal grant of the Silesian University $\mathrm{SGS/25/2024}$. A.Y. acknowledges the support from the GAUK project No. 102323 and ESA's Archival Research Visitor Programme. The research of DC has been funded by the Alexander von Humboldt Foundation and the grant ANID-FONDECYT Regular 1251444. S.R.S. acknowledges financial support from the Deutsches Zentrum für Luft- und Raumfahrt (DLR) grant FKZ 50OR2108 and support from the grant ANID FONDECYT 11240206. S.R.S. also acknowledges the support from the PID2022-136598NB-C33 grant funded by MCIN/AEI/10.13039/501100011033 and by “ERDF A way of making Europe”.\\ 
     This research has made use of data, software, and/or web tools obtained from the High Energy Astrophysics Science Archive Research Center (HEASARC), a~service of the Astrophysics Science Division at NASA/GSFC and of the Smithsonian Astrophysical Observatory’s High Energy Astrophysics Division. \\ 
     The main results of this work are based on observations obtained with the ESA science mission XMM-Newton, with instruments and contributions directly funded by ESA Member States and NASA. We also acknowledge the use of public data from the Swift data archive and Chandra Data Archive provided by the Chandra X-ray Center (CXC).\\
     The AGILE Mission is funded by the Italian Space Agency (ASI) with scientific and programmatic participation by the Italian National Institute for Astrophysics (INAF) and the Italian National Institute for Nuclear Physics (INFN). This investigation was partially supported by the ASI grant I/028/12/7-2022.

\end{acknowledgements}

%
\bibliographystyle{aa} 
\bibliography{refs} 
%

\makeatletter
\renewcommand\appendix{\par
  \setcounter{section}{0}%
  \setcounter{subsection}{0}%
  \gdef\thesection{\Alph{section}}}
\makeatother

\end{document}